\documentstyle[12pt]{article}
\begin{document}
\def\reff{\vskip 0.5cm\par\noindent\hangafter=1\hangindent=1cm} 
\baselineskip=20pt
\textwidth6.5in
\textheight8.5in
\oddsidemargin0in
\topmargin-0.25in
\pagestyle{empty}
\begin{center}
\bigskip

\rightline{FERMILAB-Pub-96/068-A}

\vspace{.2in}
{\Large \bf Comment on a Paper by
Fang, Huang, and Wu}
\bigskip

\vspace{.2in}
Scott Dodelson\\
Albert Stebbins\\
\vspace{.2in}
{\it NASA/Fermilab Astrophysics Center\\
Fermi National Accelerator Laboratory, Batavia, IL~~60510-0500}\\

\end{center}

\vspace{.3in}

\bigskip

\pagestyle{plain}

Before decoupling in the early universe, the tightly
coupled photon/electron gas underwent acoustic
oscillations. These oscillations should be visible
today in the spectrum of anisotropies. Recently 
Fang, Huang, and Wu (1996) claimed that
when random processes are accounted for, the phases
of these oscillations are no longer coherent. In
fact, they claim that the well-defined peaks will be
completely smoothed out. We show here that their claim is
incorrect. The standard Boltzmann treatment determining
the anisotropies is sufficient; random processes do
not change the standard result.

Our criticism has two parts, both dealing with their
equation
$26$. The first point is that -- at least to a first approximation
-- the power spectrum today is caused by the monopole
at decoupling. If $\Theta_0$ were zero at decoupling there
would [again to first approximation] be no anisotropy today.
Fang et al. derive an expression for the phase
shift for each of the $l-$ modes (their eqn. $25$),
but  the
only one of these phase shifts that is relevant is the $l=0$
mode. All the higher modes [with the minor exception of the 
$l=1$ mode] have negligible amplitudes in the tightly
coupled regime. The final anisotropy spectrum today depends
then only on this one phase shift $\delta \phi_0(\eta_*)$.
So we disagree with eqn $26$ which assumes that the shift in a
given multipole of the power spectrum depends on the phase shift of
that multipole at decoupling. 
To make this slightly more quantitative, let us write
the standard expression for the power spectrum:
$$ C_l = {2\over \pi} \int dk k^2 j_l^2(k\eta_0)
\left[ 
\Theta_0(k,\eta_*) + \psi(k,\eta_*) \right]^2
$$
The spherical Bessel functions take care of the free-streaming
from the last scattering surface to us today.
This equation neglects the small contributions from the dipole
and from the integrated Sachs-Wolfe effect, but this
should be irrelevant for the present purposes. If we
write $\Theta_0(\eta_*) + \psi(\eta_*) = 
A \cos\left(\phi_0(\eta_*)
+ \delta\phi_0(\eta_*) \right)$, then the change in the power spectrum
due to a change in the phases should be roughly 
$$
{\Delta C_l\over C_l}  \propto 
\delta\phi_0^2(k=l/\eta_0,\eta_*)
$$
This equation should replace equation $26$ in their paper.

Our second point is that the shift $\delta\phi_0$
is zero! The authors point this out earlier in their paper.
Physically it makes sense that $\delta \phi_0(\eta_*)=0$, for
energy must be conserved in interactions.
Since $\delta\phi_0$ is zero due to random fluctuations, the change in
the power spectrum is also zero.
There may be small residual corrections due to fluctuations in the 
quadrupole at recombination (because the photons and electrons are
tightly coupled, no momentum is lost from the dipole so $l=1$
fluctuations are also strongly suppressed)
but these should be very small. We conclude that phase randomization
does not disrupt the structure of the ``Doppler peaks.''

We thank Lam Hui for a helpful comment. 
This work was supported in part by the DOE 
and NASA (at Fermilab through grant NAG 5-2788).

\begin{center}
{\bf References}
\end{center}

\reff Li-Zhi Fang , Zheng Huang , $\&$ Xian-Ping Wu 1996,
astro-ph 9601087

\end{document}